\documentclass[]{spie} 
\usepackage{amsmath}  
\usepackage{graphicx}
\usepackage[breaklinks,colorlinks,urlcolor=blue,linkcolor=blue,citecolor=blue]{hyperref}
\usepackage{longtable}
\usepackage{color}

\title{LAMP: a micro-satellite based soft X-ray polarimeter for astrophysics}
\author{Rui She\supit{1},
Hua Feng\supit{1},
Fabio Muleri\supit{2},
Paolo Soffitta\supit{2},
Renxin Xu\supit{3},
Hong Li\supit{1},
Ronaldo Bellazzini\supit{4},
Zhanshan Wang\supit{5},
Daniele Spiga\supit{6},
Massimo Minuti\supit{4},
Alessandro Brez\supit{4},
Gloria Spandre\supit{4},
Michele Pinchera\supit{4},
Carmelo Sgr\`{o}\supit{4},
Luca Baldini\supit{4},
Mingwu Wen\supit{5},
Zhengxiang Shen\supit{5},
Giovanni Pareschi\supit{6},
Gianpiero Tagliaferri\supit{6}, 
Kashmira Tayabaly\supit{6}, 
Bianca Salmaso\supit{6}, and
Yafeng Zhan\supit{7}
\skiplinehalf
\supit{1}Department of Engineering Physics and Center for Astrophysics, Tsinghua University, Beijing 100084, China; \\
\supit{2}IAPS/INAF, Via Fosso del Cavaliere 100, Rome 00133, Italy; \\
\supit{3}Kavli Institute for Astronomy and Astrophysics, Peking University, Beijing 100871, China; \\
\supit{4}INFN-Pisa, Largo B. Pontecorvo 3, Pisa 56127, Italy; \\
\supit{5}Institute of Precision Optical Engineering, Tongji University,  1239 Siping Road, Shanghai 200092, China; \\
\supit{6}INAF/Osservatorio Astronomico di Brera, Via Bianchi 46, Merate 23807, Italy;\\
\supit{7}Space Center, Tsinghua University, Beijing 100084, China}

\begin{document}
\maketitle

\begin{abstract}
The Lightweight Asymmetry and Magnetism Probe (LAMP) is a micro-satellite mission concept dedicated for astronomical X-ray polarimetry and is currently under early phase study. It consists of segmented paraboloidal multilayer mirrors with a collecting area of about 1300 cm$^2$ to reflect and focus 250 eV X-rays, which will be detected by position sensitive detectors at the focal plane. The primary targets of LAMP include the thermal emission from the surface of pulsars and synchrotron emission produced by relativistic jets in blazars. With the expected sensitivity, it will allow us to detect polarization or place a tight upper limit for about 10 pulsars and 20 blazars. In addition to measuring magnetic structures in these objects, LAMP will also enable us to discover bare quark stars if they exist, whose thermal emission is expected to be zero polarized, while the thermal emission from neutron stars is believed to be highly polarized due to plasma polarization and the quantum electrodynamics (QED) effect. Here we present an overview of the mission concept, its science objectives and simulated observational results.
\end{abstract}

\section{Introduction}
Astronomical X-ray polarimetry in the keV band has been an unexplored area for 40 years, since the last successful detection in 1975 with a Bragg polarimeter onboard the OSO-8 satellite \cite{Weisskopf1976a}. The polarimeter consists of small graphite crystals mounted on two segments of paraboloidal surface, which reflect 2.6 keV  (or 5.2 keV on the 2nd order) X-rays near 45 degrees and steer them onto the gas proportional counters.  Rotation of the instrument along its optical axis produces azimuthal modulation of the reflected flux, if it is partially or fully polarized, as the reflectivity of Bragg diffraction depends on the polarization angle of incident photons. Bragg crystals tilted at 45 degrees can act as a perfect X-ray polarization analyzer. However, the narrow bandwidth of the Bragg reflectors, even with mosaic and imperfect crystals that result in a relatively wider passband, limits the sensitivity in polarimetry.  With the instrument on OSO-8, they detected a linear polarization of 19.2\% $\pm$ 1.0\% at a position angle of $156.4^\circ \pm 1.4^\circ$ for the Crab nebula\cite{Weisskopf1976b,Weisskopf1978a}, and placed a time-averaged upper limit of 1--3 per cent for Sco X-1 \cite{Weisskopf1978b,Long1979} or a few per cent for Cyg X-1 and X-2 \cite{Long1980,Hughes1984}. While for other sources, only loose upper limits of 10\% or above could be obtained, which are not sufficiently useful to constrain the physical models\cite{Hughes1984}.  

The photoelectric effect is the dominant mechanism for keV X-ray photons interacting with matter.  The emission of photoelectrons is not isotropic but is most likely along the direction of the electric vector of the incident photon, which allows us to reconstruct the polarization degree and angle of the X-ray beam.  In 2001,  it is first demonstrated that sensitive X-ray polarimetry in the 2-10 keV band can be realized via measuring 2-D images of photoelectrons in the gas using micro-pattern gas detectors \cite{Costa2001,Bellazzini2003,Bellazzini2003b,Bellazzini2006a,Bellazzini2007a}.  On the basis of such techniques, future X-ray telescopes with polarimetry capability were proposed and under development \cite{Jahoda2010,Soffitta2013}.

The invention of photoelectric polarimeters will make a breakthrough in astronomy as it increases the sensitivity of X-ray polarimetry by orders of magnitudes in the energy band of 2-30 keV. However, it is hard to measure polarization for photons below 2 keV due to two reasons: the spatial resolution of gas detectors of about 100 $\mu$m is insufficient to resolve tracks of low energy electrons, and the Auger electrons of comparable energies will confuse the identification of photoelectrons. 

So far, the best approach for X-ray polarimetry below 2 keV is still Bragg diffraction at 45 degrees, but with multilayer coated mirrors instead. Compared with natural crystals, multilayer mirrors offer us two advantages. It allows us to choose, to some extent,  the energy of interest, and the mirror can be shaped to a nearly perfect paraboloidal surface to focus the incident beam onto a tiny spot such that the signal to noise ratio is maximized. A multilayer mirror based Bragg polarimeter was first proposed with a configuration similar to the one on OSO-8 \cite{Marshall2003}. An early version of the GEMS mission concept also included a small piece of thin mirror in the light path above the focal plane to reflect soft X-ray at 0.5 keV for polarization analysis along with the spinning of the telescope \cite{Allured2012}.  It is also suggested that a multilayer mirror with gradient thickness associated with an X-ray grating can act as a broadband soft X-ray polarimeter \cite{Marshall2007}. 

Here, following the configuration of the paraboloidal Bragg polarimeters, we propose a concept for a soft X-ray polarimeter, the Lightweight Asymmetry and Magnetism Probe (LAMP), which measures the polarization of monochromatic X-rays near 250~eV and can fit in a micro-satellite. Its compactness and technical readiness give rise to a low cost and high feasibility. We argue that one is able to perform unique, interesting science cases with LAMP, in addition to adding a low energy point at 250 eV for the broadband photoelectric polarimeter that works above 2 keV. 

\section{Science objectives with LAMP}

\subsection{Thermal emission from the surface of pulsars}

One of the major puzzles in high energy astrophysics is related to the nature of neutron stars, where matters are compressed to super-nuclear densities.  The interior of neutron stars is poorly understood,  i.e., what is the state of matters under the extreme density \cite{Lattimer2001}. Thermal emission in the soft X-ray band with a temperature on the order of $10^2$ eV has been detected from different populations of neutron stars \cite{Potekhin2014}.  The thermal emission is thought to originate from the atmosphere of the neutron star, a thin layer of plasma which is strongly magnetized. It is argued that such emission may offer us a useful probe to the magnetic fields and interior physics of neutron stars \cite{Lai2010}. As electrons cannot move freely in the direction perpendicular to the magnetic field, the Thomson scattering cross-section for a soft X-ray in magnetized plasma is a function of its polarization angle. Under normal physical conditions, the optical depth for photons in the extraordinary mode (X-mode), whose polarization vector is perpendicular to the magnetic field, is greatly suppressed compared with photons in the ordinary mode (O-mode), whose polarization is in line with the magnetic field. As a result, X-mode photons arise from a photosphere that is deeper and hotter with a higher flux than for O-mode photons in the neutron star atmosphere, and on each local region on the surface, the thermal emission is highly polarized.  When the photon leaves the neutron star surface and propagates in the magnetosphere, due to the quantum electrodynamics (QED) effect that the vacuum is birefringent, the two modes are decoupled and the polarization will rotate and follow the magnetic field adiabatically until a large distance, the polarization-limiting radius, where the two modes couple again \cite{Heyl2000,Heyl2002}.  This effect leads to a consequence that the net observed polarization is not directly summed from photons on the surface, rather it is determined by the average field orientation at the polarization-limiting radius.  Because the fields at large radii along the line of sight are much more ordered than on the surface, the QED vacuum birefringence gives rise to a large net polarization for the thermal emission.  

Thus, phase resolved polarimetry for the thermal emission will allow us to measure the geometry of neutron star magnetic fields. It is only sensitive to the dipole component as it can extend to large radii. Plus, measurement of a high degree of polarization will be a strong evidence for the QED effect, which cannot be tested in the lab by far. For such measurements, the best targets are the X-ray dim isolated neutron stars (XDINSs) discovered by ROSAT, which show pure thermal emission in the X-ray band and there is no blending with the non-thermal component. 

One major debate is that whether (or which) pulsars are neutron stars or quark/quark-cluster stars.  Unlike neutron stars which are gravity-bound, quark stars are self-bound so that they do not need an atmosphere (a pressure gradient) to link the stellar body and the vacuum. Most of the quark star models suggest that they are bare quark stars.  The high thermal conductivity on the surface of quark stars due to degenerated electrons indicates that photons of the two modes will have the same temperature. Therefore, thermal emission from the surface of bare quark stars will be almost zero polarized \cite{Lu2013}, and soft X-ray polarimetry will enable us to discover bare quark stars if they exist. 

\subsection{Relativistic jets in blazars}

It is still unknown how the relativistic jets are launched around accreting black holes. It is believed that the magnetic fields play a key role in the formation and collimation of the jets, via the rotation of the black hole or the accretion disk \cite{BlandfordZnajek1977,BlandfordPayne1982}. Thus, it is of great importance to measure the structure of the magnetic fields at the base of the relativistic jets where the matter gets accelerated. Radiation from blazars is ideal for such measurements because the jet emission is a dominant component over the disk emission in most cases due to Doppler boosting. We will choose the high synchrotron peak (HSP) blazars whose synchrotron emission peaks in the X-ray band, as the synchrotron radiation is highly polarized and its position angle is directly related to the orientation of the local magnetic fields.

Optical observations have demonstrated that polarimetry for synchrotron radiation from the blazars is capable of revealing the geometry of the magnetic fields in the jets. Optical polarimetry for BL Lac led to the discovery of a continuous change of the polarization angle by $\sim$200 degrees during a flare \cite{Marscher2008}. Later on, a smooth variation of the position angle around 600 degrees and 180 degrees were detected for PKS 1510-089 \cite{Marscher2010} and 3C 279 \cite{Abdo2010}, respectively.  These were interpreted as due to the propagation of shocks in helical magnetic fields. X-ray polarimetry may do a better job, because the X-ray emission region is thought to be more compact, allowing us to probe the local field rather than an averaged effect. The HSP blazars like Mrk 421 are good candidates for such measurement. 

\subsection{Other high energy objects}

The two kinds of sources mentioned above will be the major scientific goals for LAMP, due to their brightness in the soft X-ray band (or specifically at 250 eV) and high degrees of polarization.  For other types of high energy sources, LAMP can offer a low energy measurement in addition to the photoelectric polarimeters. As the polarization is a function of energy, an additional point at 250 eV which is one order of magnitude below the lower threshold ($\sim2$ keV) for photoelectric polarimeters, may provide useful constraints onto the geometry or the magnetic fields of the source.  

X-ray polarimetry can help disentangle the coupling between the black hole spin and inclination, providing an unbiased estimate of the black hole spin via spectral fitting in the thermal state \cite{Li2009,Schnittman2009}. It can help test the geometry for active galactic nuclei (AGNs) and the unification model \cite{Goosmann2011}.  With X-ray polarimetry, one is able to distinguish competing models for the high energy emission from rotation-powered pulsars \cite{Harding2004}, magnetized accretion-powered pulsars \cite{Meszaros1988}, and millisecond accreting pulsars \cite{Viironen2004}. For extended sources like the pulsar wind nebulae, X-ray polarimetry can tell us the information about the magnetic fields and distinguish between synchrotron and bremsstrahlung emission, but they are in fact better targets for imaging polarimeters like XIPE as LAMP cannot spatially resolve them. 

\section{Instrument}

\begin{figure}[htbp]
  \begin{center}
	\includegraphics[width=0.6\linewidth]{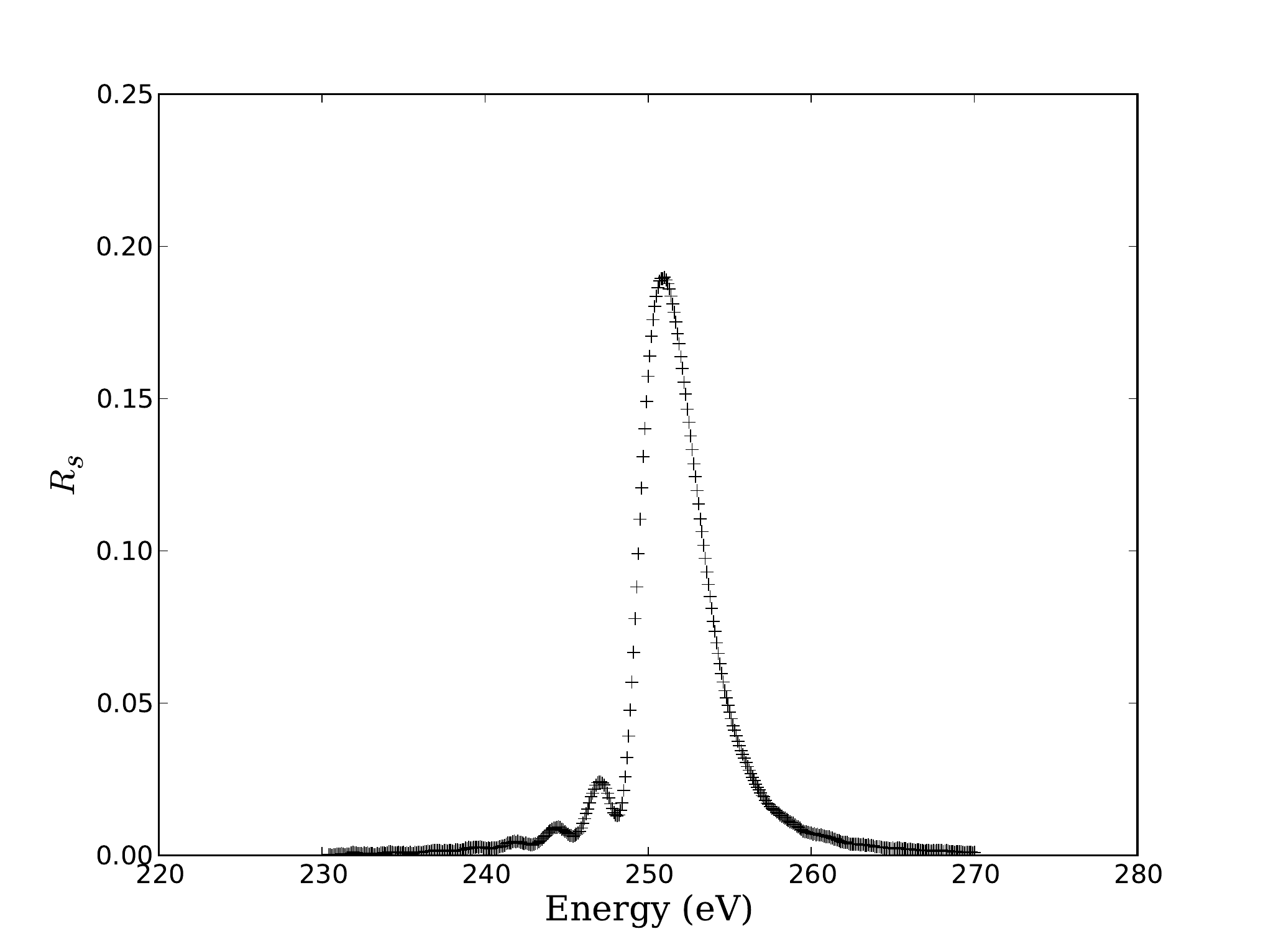}
  \end{center}
  \caption{Measured reflectivity curve at an incident angle of 45$^{\circ}$.}
  \label{fig:ref_curve}
\end{figure}

The science payload of LAMP consists of the optics and the detectors. The optics is a combination of 16 segments of paraboloidal mirrors with multilayer coating. The reflection angle varies from 50 degrees at the upper edge to 40 degrees at the bottom edge of the mirrors. A gradient change on the thickness of the multilayer along the radius is required, to satisfy the Bragg law for 250 eV photons anywhere on the mirror. An accurate control of the multilayer thickness is needed but not critical, as the throughput is the most important parameter while the monochromaticity is not a requirement because the energy resolution of the detectors is much larger than the bandwidth of the reflectivity. The reflectivity for s-light is a function of the photon energy, $R(E)$, and we are pursuing a high throughput for photons at all energies, that is, to maximize the effective bandwidth $\int{R(E)dE}$. There are two viable choices for the base of the mirror, electroformed nickels or slumped glasses. Both materials can be shaped to a paraboloidal surface against a mandrel. These techniques are under investigation at the Tongji University and the Osservatorio Astronomico di Brera.  The choice for the multilayer materials is also under study. So far, a Co/C coated mirror on a flat glass sample was tested to have a peak reflectivity for 250 eV s-light of nearly 0.2 and an energy-integrated reflectivity of 0.97 eV at an incident angle of 45 degrees (see Figure~\ref{fig:ref_curve}; also see Ref.~\citenum{Spiga15} in the same volume of proceedings). The geometric collection area is around 1300 cm$^2$. Each mirror segment will focus the incident beam into a single focal spot on the focal plane. Four detectors, each facing four segments of mirrors, will measure the azimuthal distribution of the reflected fluxes. 

\begin{figure}[htbp]
  \begin{center}
	\includegraphics[width=0.42\linewidth]{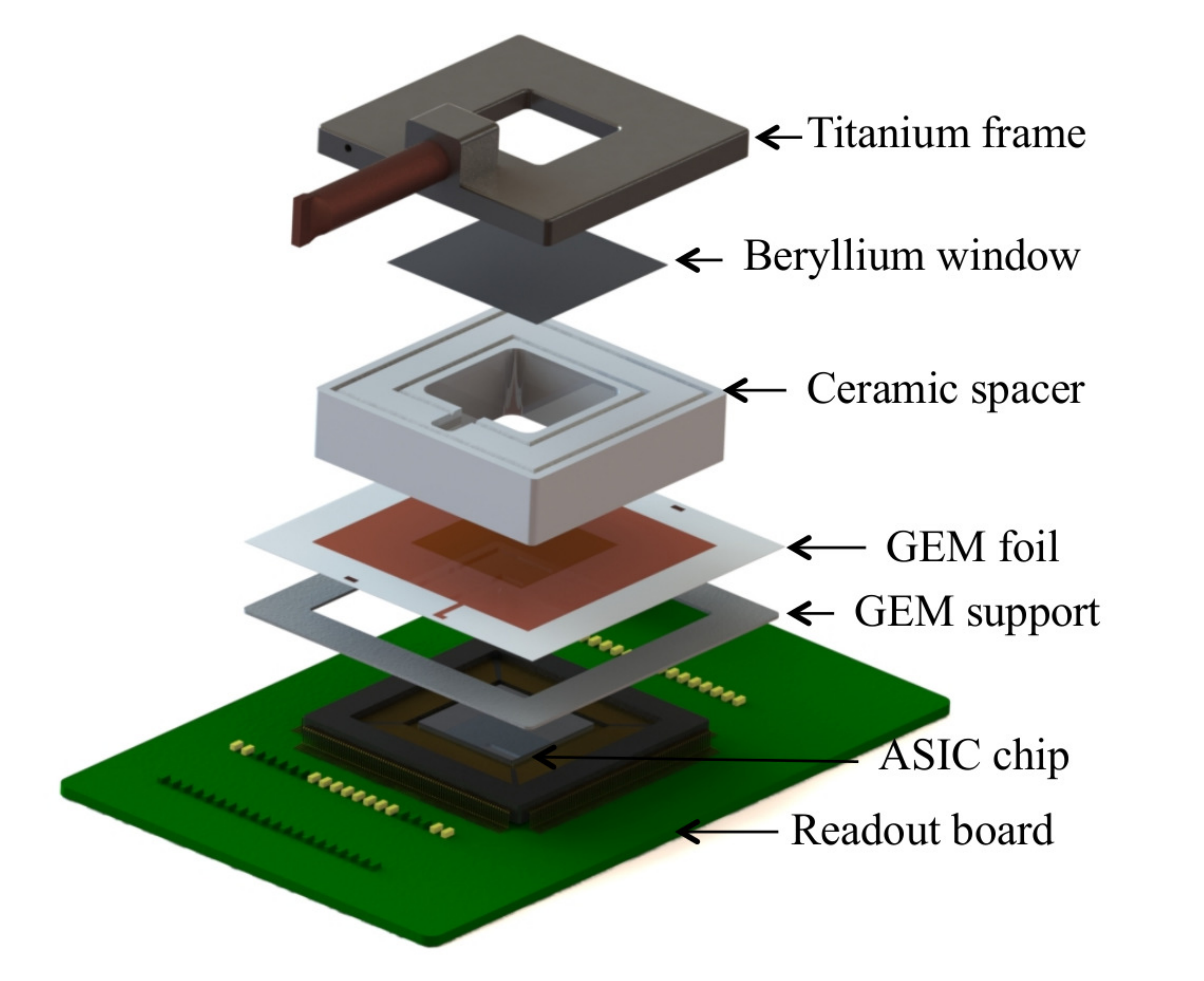}
	\includegraphics[width=0.53\linewidth]{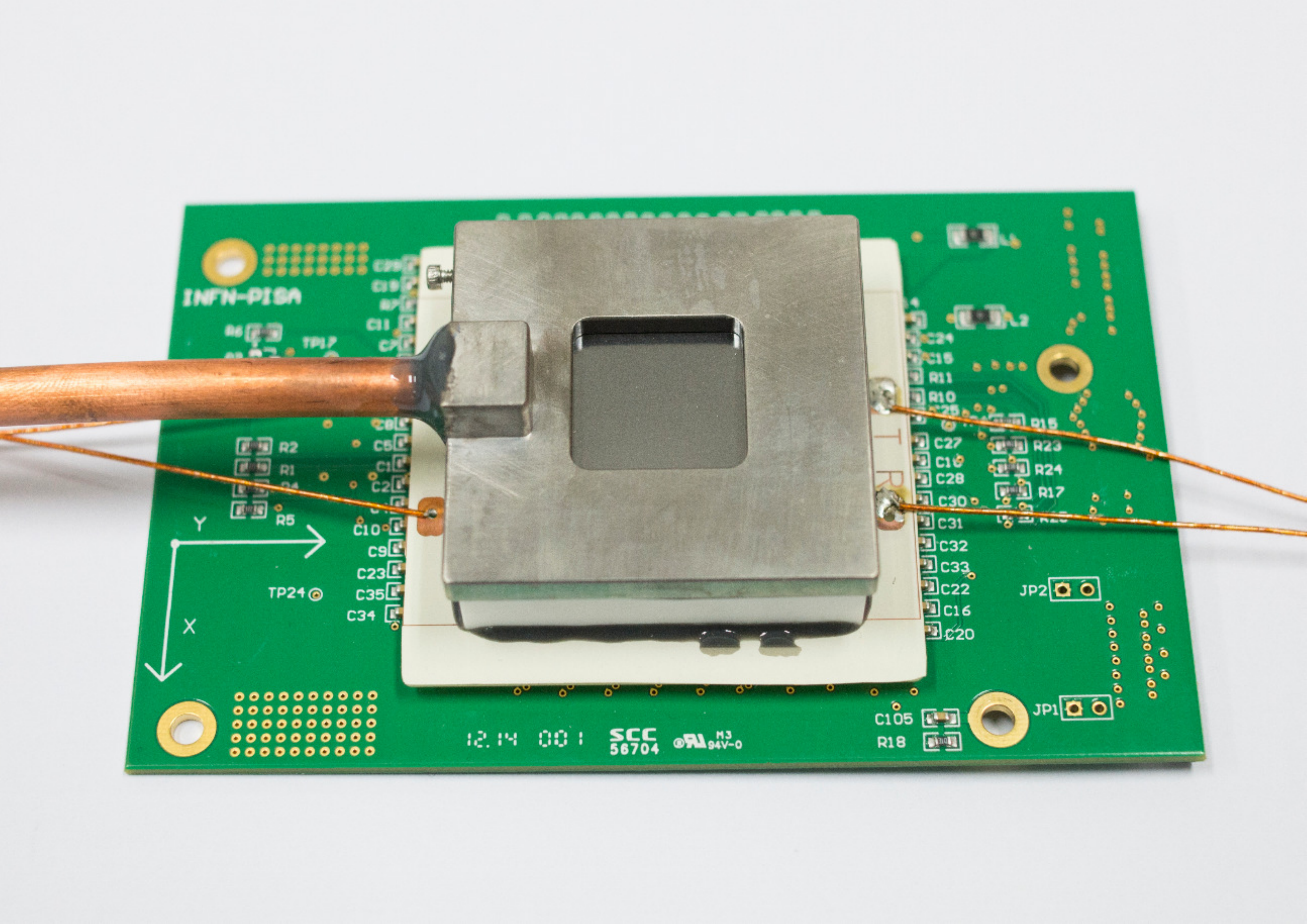}
  \end{center}
  \caption{The structure of the GPD detector (left) and the one assembled at Tsinghua University (right).}
  \label{fig:gpd}
\end{figure}

\begin{figure}[tbp]
  \begin{center}
	\includegraphics[width=0.6\linewidth]{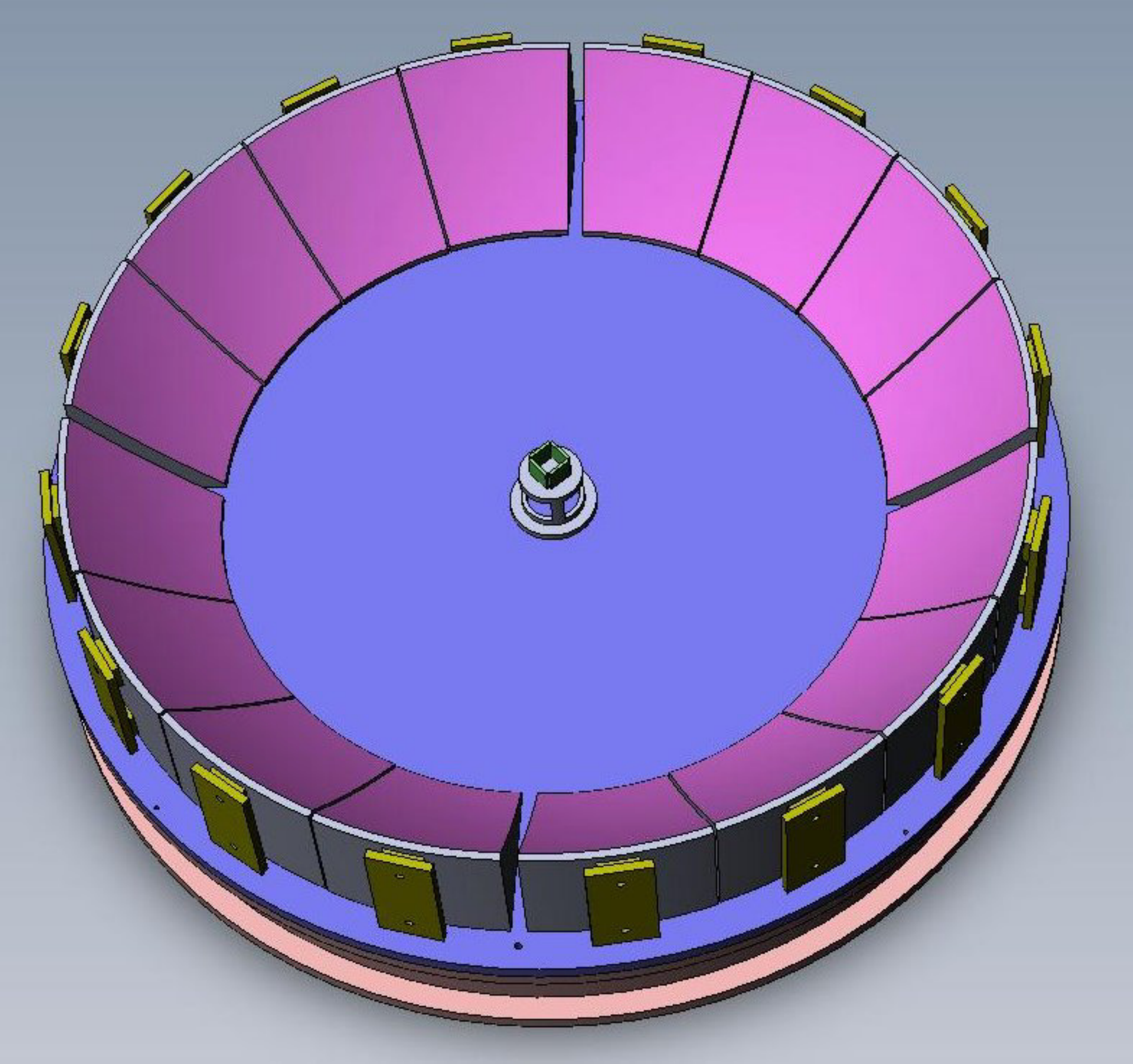}
  \end{center}
  \caption{A schematic drawing of the LAMP instrument.}
  \label{fig:mirror}
\end{figure}

\begin{figure}[htbp]
  \begin{center}
	\includegraphics[width=0.6\linewidth]{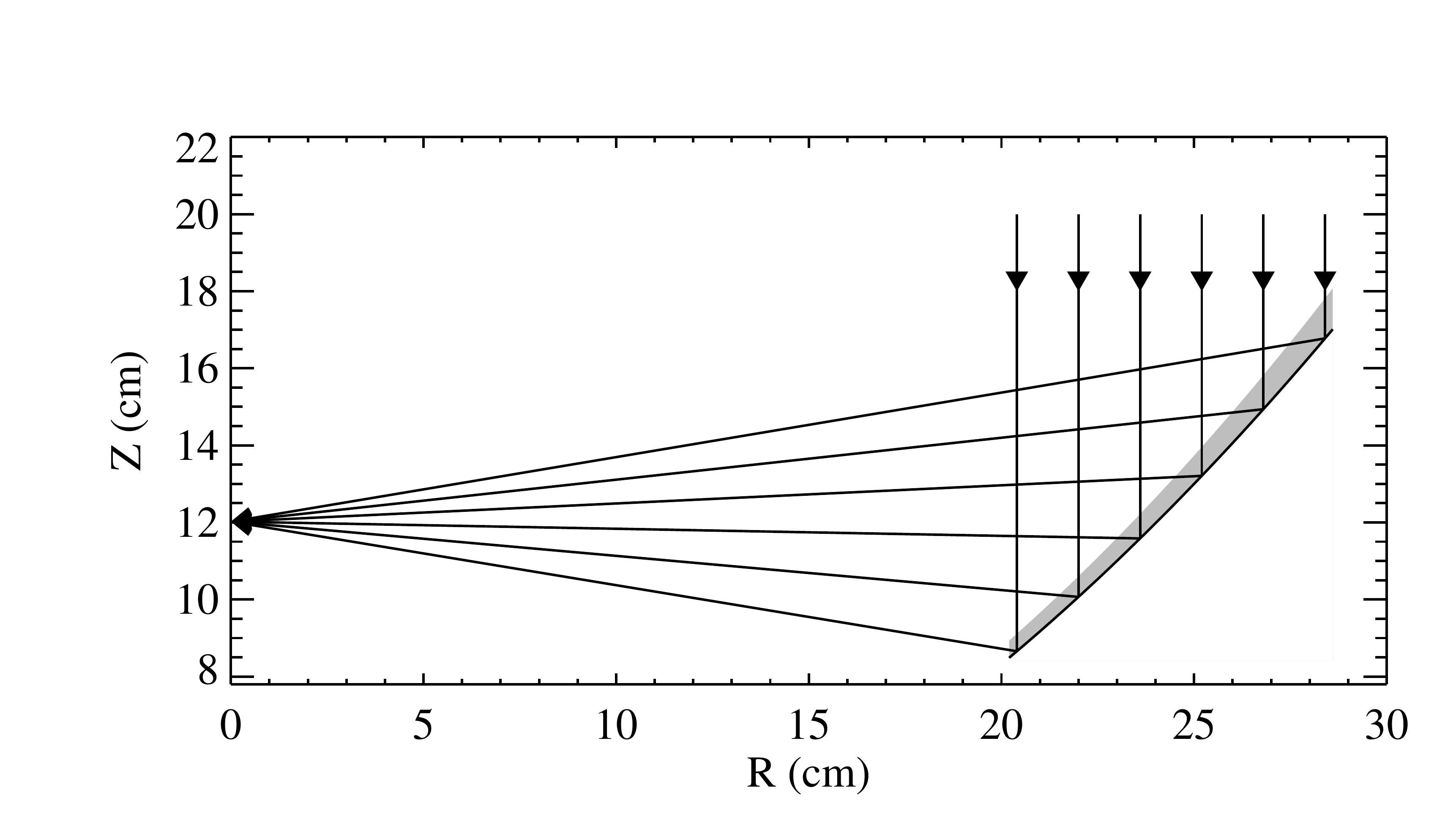}
  \end{center}
  \caption{Radial cross-section of the mirror and light paths. The thickness of each layer is larger at the upper end and smaller at the lower end, in order to reflect and focus monochromatic 250 eV photons.}
  \label{fig:path}
\end{figure}

There are several choices for the focal plane detectors. Semiconductor detectors like CCDs may be ideal choices for this purpose due to their high position and spectral resolution. However, for a low-cost micro-satellite,  it is difficult to implement CCDs which require a lot of resources for cooling and protection from radiation damage. Therefore, we choose to use gas detectors, which work at room temperature, with an ultrathin window for imaging. The ultrathin window developed by HS Foils Oy has a transmission of $\sim$35\% near 250 eV. The gas pixel detector (GPD) developed by INFN-Pisa \cite{Bellazzini2006a,Bellazzini2006b,Bellazzini2006c,Bellazzini2007a,Bellazzini2007b,Bellazzini2007c,Muleri2009,Muleri2010,Bellazzini2013} for photoelectric polarimetry  is just suitable for this application (see Figure~\ref{fig:gpd}). The GPD detector has a sensitive area of 1.5cm $\times$ 1.5cm and a position resolution of less than 100$\mu$m, able to contain and resolve 4 focal spots. The gas mixture could be Ar/CO$_2$ at one atmosphere with an absorption depth of 1 cm, giving rise to a stopping power of almost 100 per cent. A schematic drawing of the LAMP payload is shown in Figure~\ref{fig:mirror}. The geometry for the mirror cross-section and photon paths is shown in Figure~\ref{fig:path}.
To smooth out the potential systematics caused by variations in reflectivity for each mirror segment and in detection efficiency for each focal spot, the satellite will spin around its optical axis slowly and constantly during observations. 

\section{Raytracing simulation}

\begin{figure}[htpb]
  \begin{center}
	\includegraphics[width=0.48\linewidth]{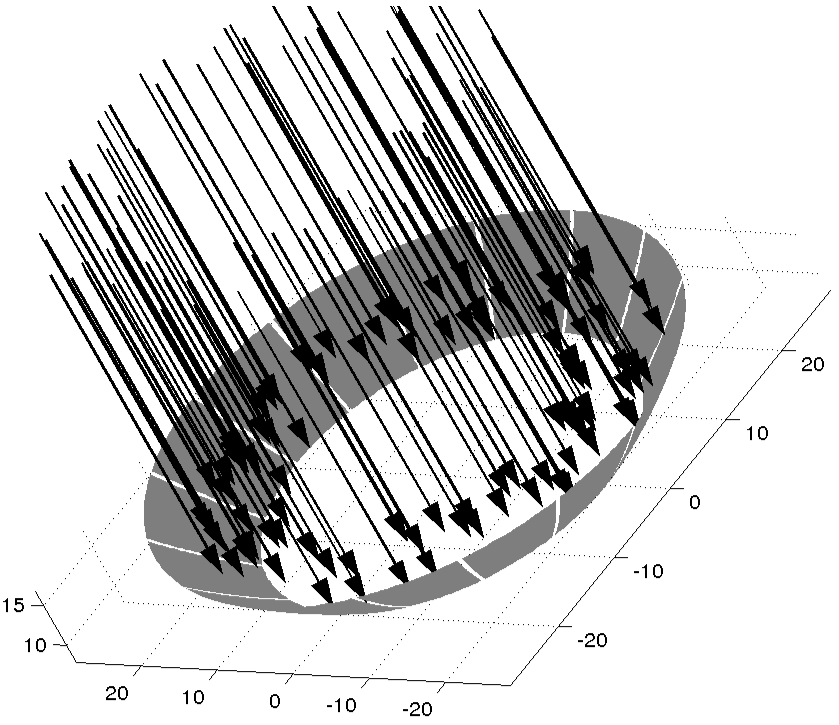}
	\includegraphics[width=0.48\linewidth]{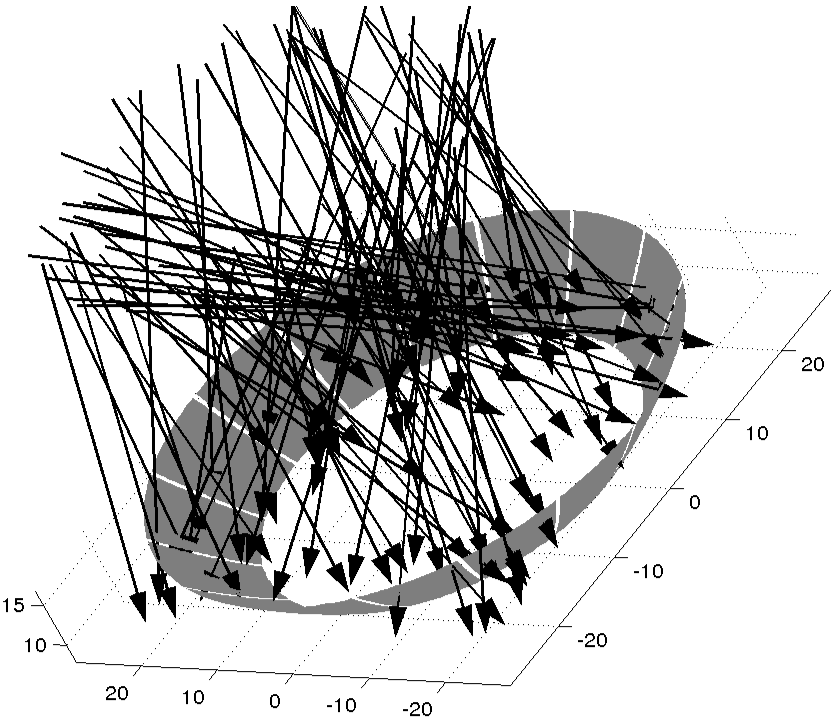}
  \end{center}
  \caption{Incident photons from a point-like source (left) and the soft X-ray diffuse background (right) in the simulation package.}
  \label{fig:photons}
\end{figure}

We ran raytracing simulations to estimate the performance of the instrument and the sensitivity in polarization. We considered photons from a point-like source in an energy range of 245--255 eV, outside of which the reflectivity is extremely low.  As the energy band is so narrow, a flat photon spectrum is assumed for simplification.  The soft X-ray diffuse background  at 1/4 keV is the main component of the background \cite{Snowden1997}. The particle background is estimated to be minor and negligible compared with the photon background; the detector is also capable of discriminating particle events on the basis of their long tracks. In the simulation, we adopted a background level of 840~photons~cm$^{-2}$~s$^{-1}$~keV$^{-1}$~sr$^{-1}$ at 250 eV based on XMM-Newton observations of blank sky regions \cite{Lumb2002}. On large scales, the background distribution is not isotropic and varies significantly at different directions. We note that this is not a problem because LAMP will only aim at bright sources in which cases the background can be ignored. Figure~\ref{fig:photons} illustrates the incident source and background photon tracks. 

Given a local position on the surface of the mirror, photons of different energies from different directions may come. The reflectivity is a function of the incident angle, photon energy, and layer thickness (or the position on the mirror).  We built a library of reflectivity in the space occupied by above parameters using the IMD package.  In the simulation, the reflectivity for every single event is interpolated from the library. 

One of the important practical issues is that the optical axis of the instrument may be misaligned with the source direction due to pointing inaccuracy and attitude instability. As a result, the photons from a point-like source will be off-axis. For an on-axis source, the focal spot is a point assuming a perfect mirror and detector.  The focal spot will rapidly grow in size along with the increase of the off-axis angle. This will dilute the image and decrease the signal to noise ratio and such effect has to be taken into account. The off-axis angle is unlikely to be a constant and will change with time.  As the pointing stability is usually much better than the pointing accuracy, we assume a constant off-axis angle during an observation.  

A second effect from an off-axis pointing is that the incident angle for each mirror segment differs from each other and so does the reflectivity, which will lead to a systematic error
in the polarization measurement. A slow rotation of the instrument around its optical axis
can vanish this effect. 

\begin{figure}[htbp]
  \begin{center}
	\includegraphics[width=0.48\linewidth]{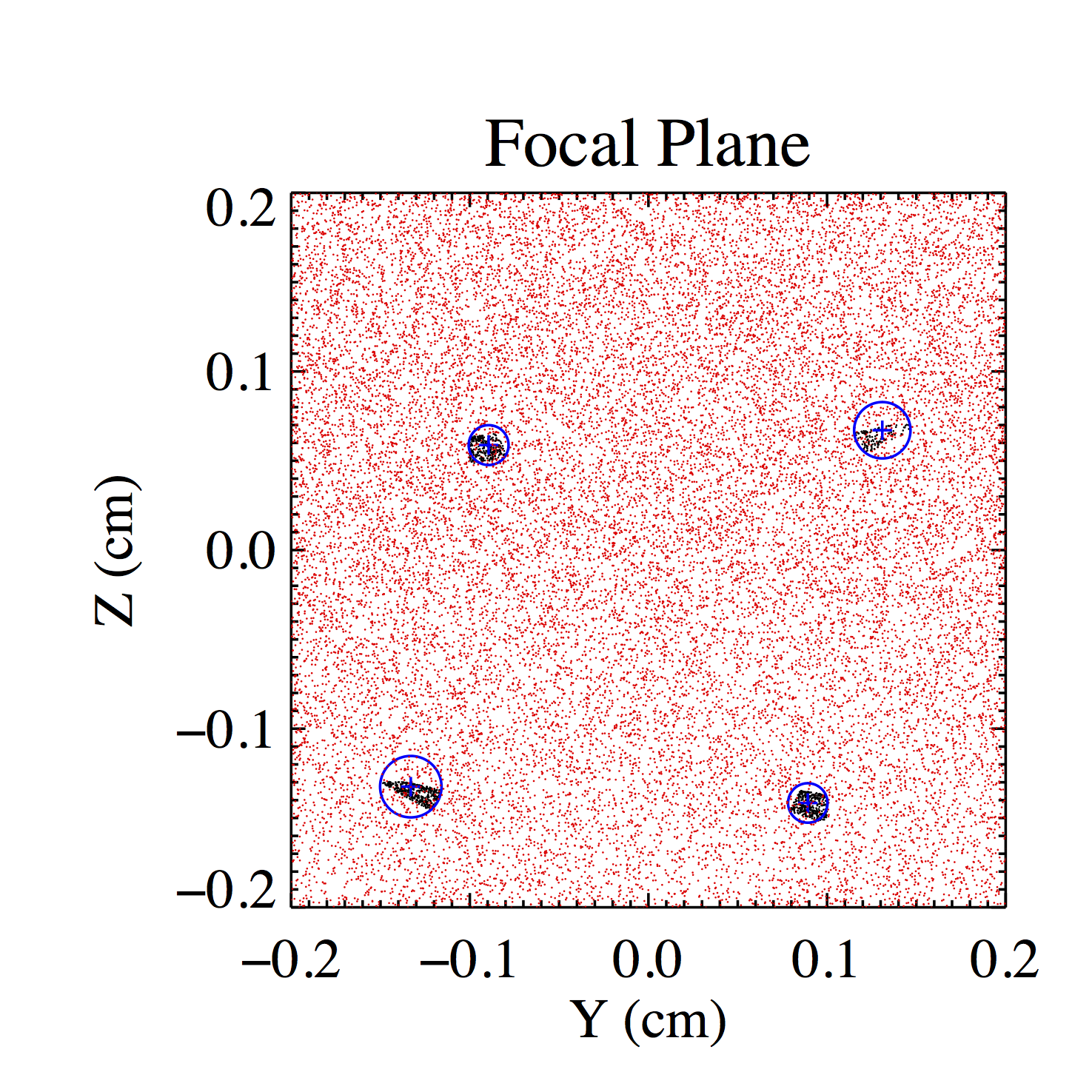}
  \end{center}
  \caption{Simulated image on the focal plane assuming an off-axis angle of 6$^{\prime}$. The red dots are the background photons and the black ones are source photons. The circles indicate the source extraction region.}
  \label{fig:bkg_offaxis}
\end{figure}

With an off-axis angle of 6$^{\prime}$, the focal spot size varies from 220~$\mu$m to 360~$\mu$m, as the focal length is the same for the mirror segments but the incident angle for the focused beam to the detector is different.  The simulated image including both the source and the background is displayed in Figure~\ref{fig:bkg_offaxis}.

\begin{figure}[htpb]
  \begin{center}
	\includegraphics[width=0.48\linewidth]{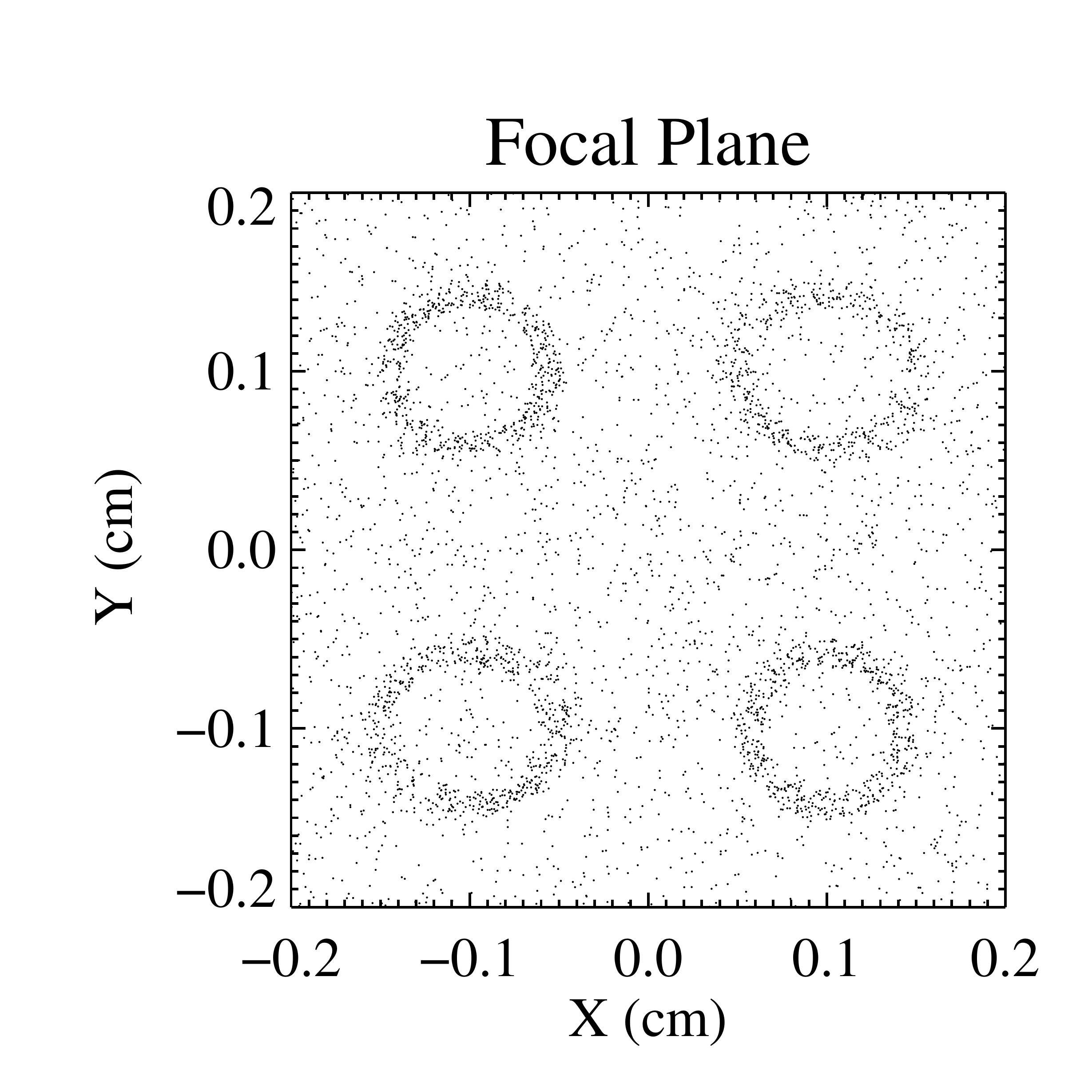}
	\includegraphics[width=0.48\linewidth]{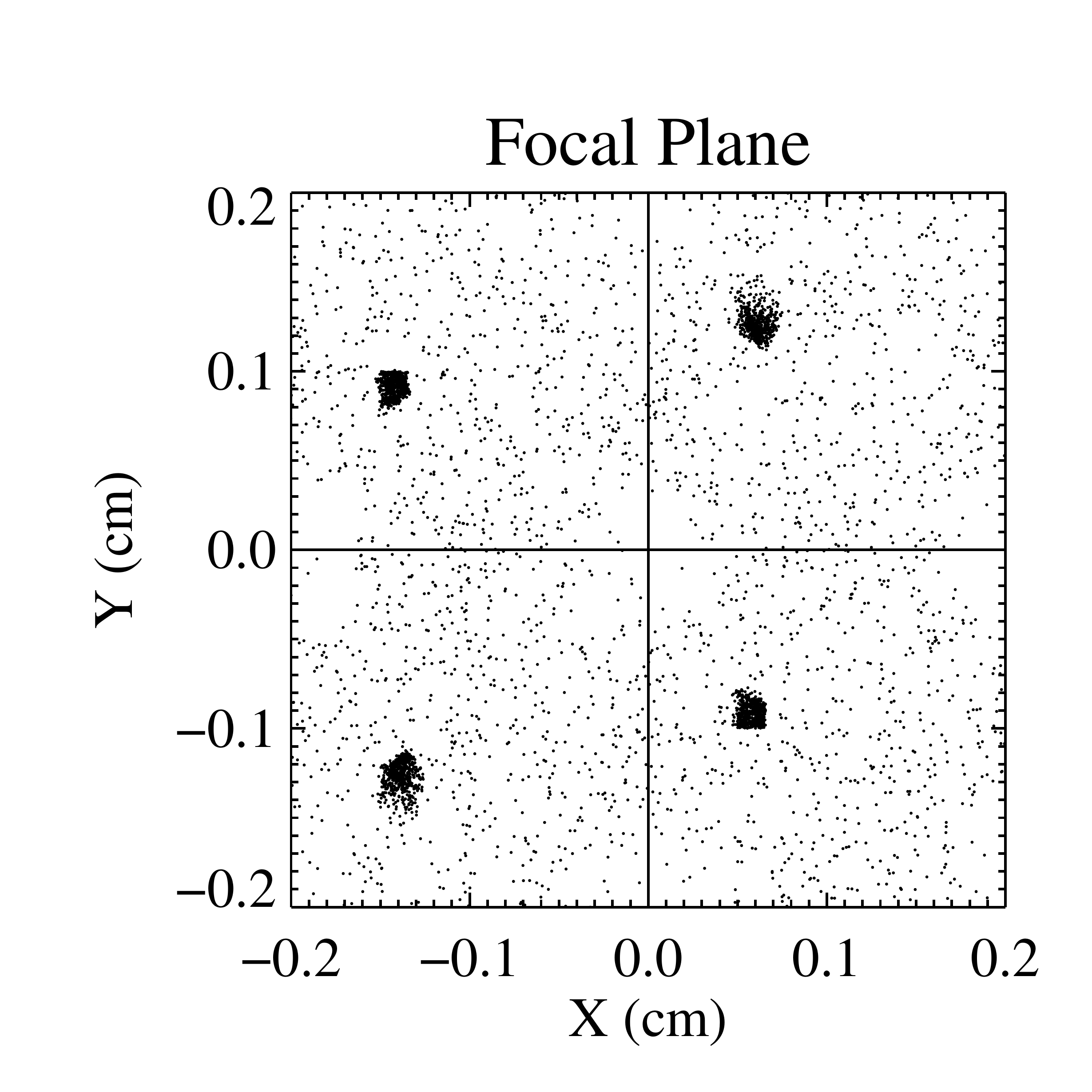}
  \end{center}
  \caption{Restoring images with rotation. Shown in left panel focus spot with
  rotation (annulus indeed), right panel is focus spot after restoring using
  satellite position records.}
  \label{fig:restore_rotation}
\end{figure}

When the instrument spins, as shown in Figure~\ref{fig:restore_rotation}, the distribution of source photons on the detector plane will become circular around the optical axis. The measurement accuracy of the pointing is usually good enough and this information will allow us to redistribute all the detected photons to the case as if there is no rotation for the instrument. Source photons will be restored back to a single spot, while background photons, which are almost evenly populated on the detector plane, are still following a random distribution.  

Also, we assumed that the instrument response is uneven for each mirror segment and detector.  A random multiplicative factor between 0.9 and 1.1 was assigned to the product of the reflectivity and the detection efficiency for each single beam. This effect will cause a systematic error because an unpolarized source will result in some level of azimuthal modulation in flux and appear like a partially polarized source. 

\begin{figure}[htpb]
  \begin{center}
	\includegraphics[width=0.48\linewidth]{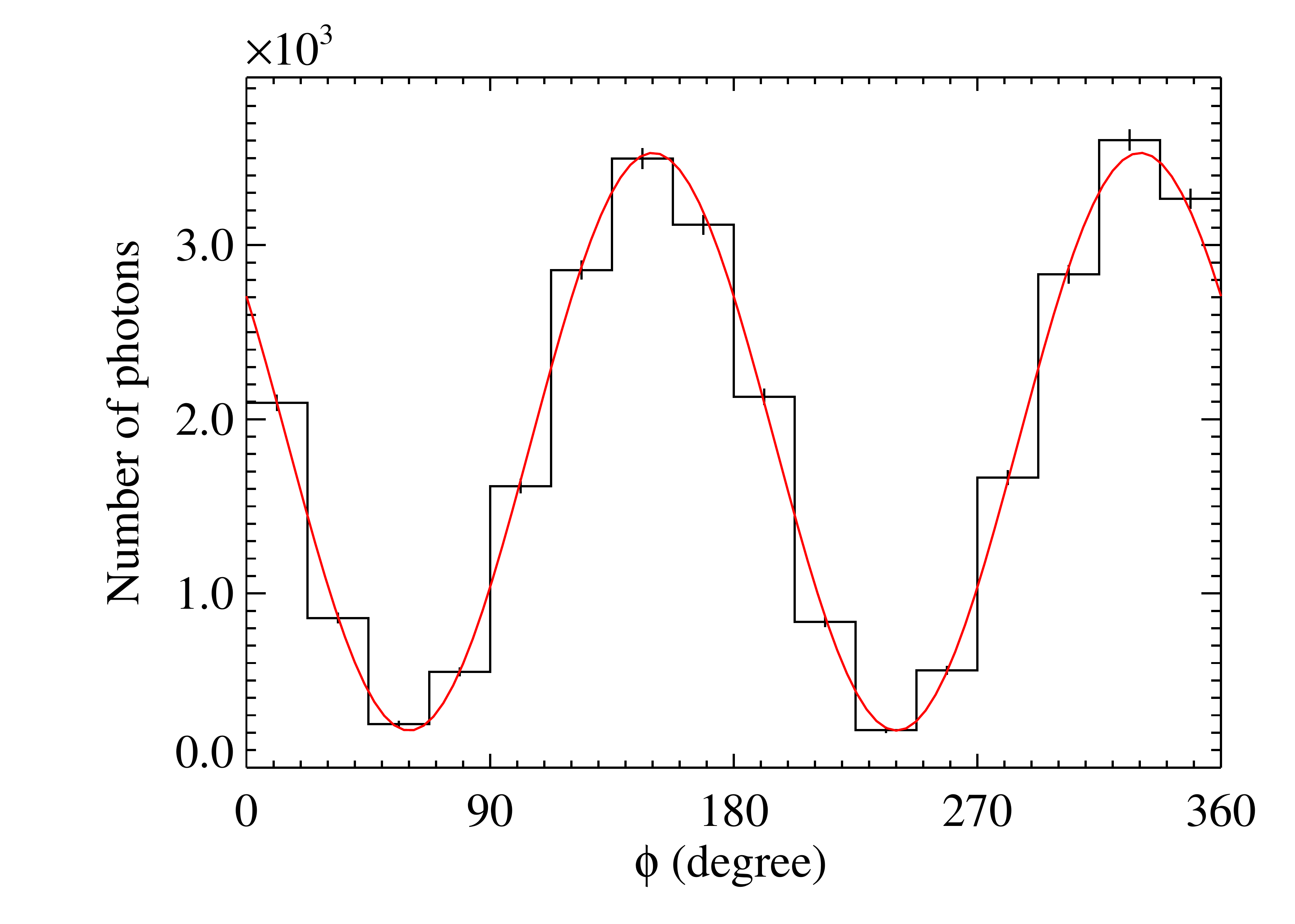}
	\includegraphics[width=0.48\linewidth]{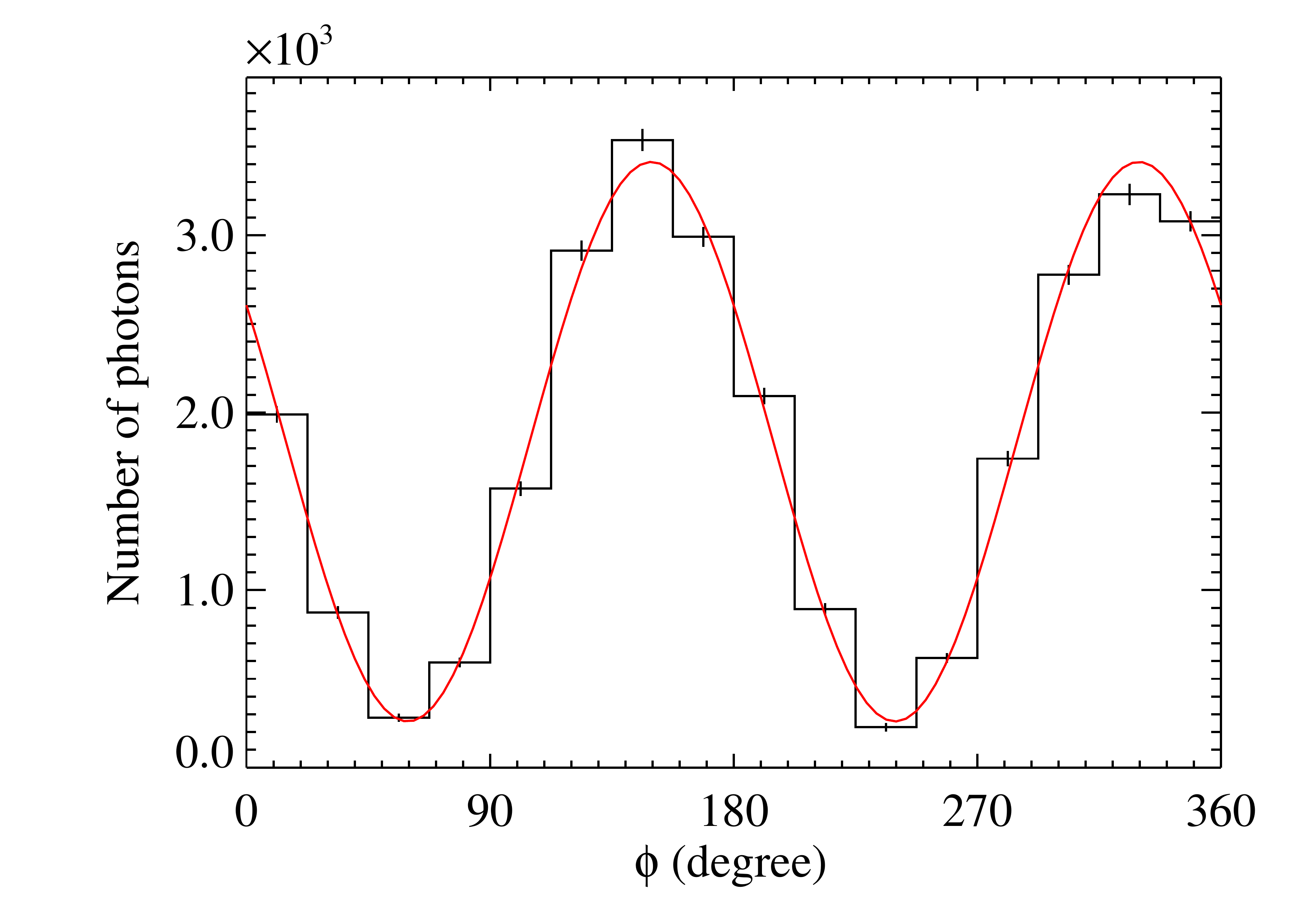}
	\includegraphics[width=0.48\linewidth]{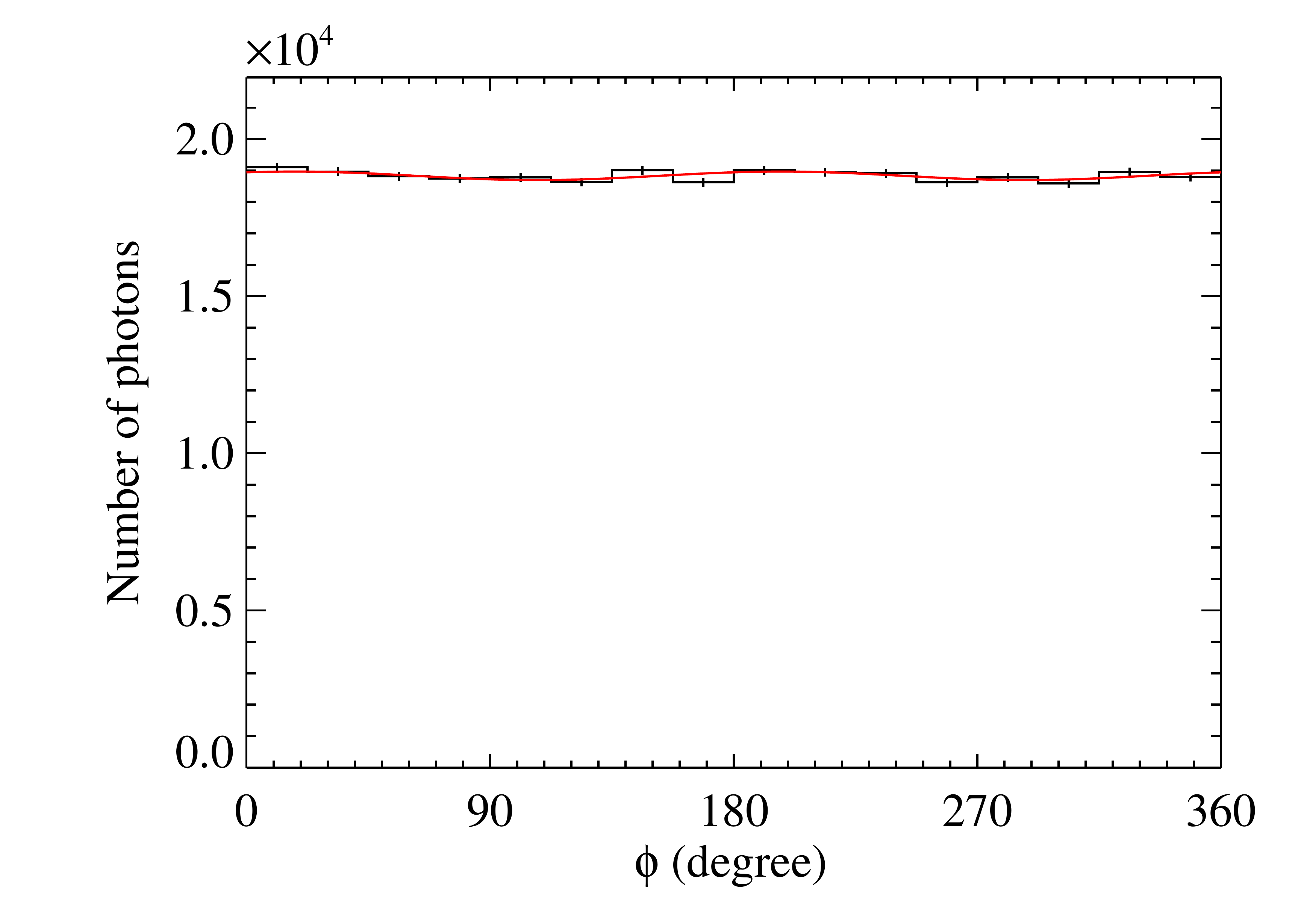}
	\includegraphics[width=0.48\linewidth]{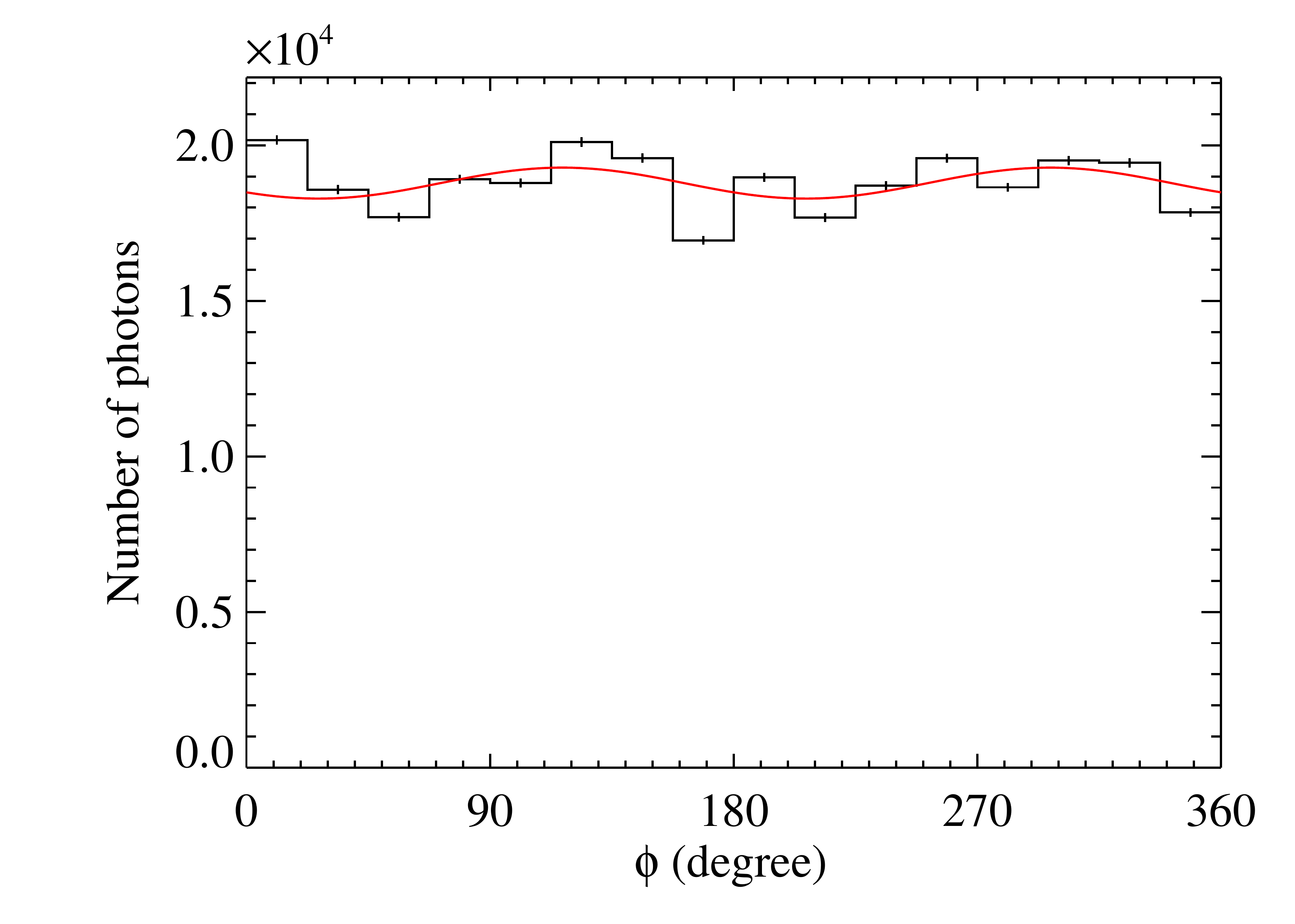}
  \end{center}
  \caption{Simulated modulation curves for a fully polarized source (top) and unpolarized source (bottom)  as bright as RX J1856.5$-$3754. The left panels are curves with instrument spinning, and the right panels are curves without spinning. For a fully polarized source, the modulation factor is $88.7\%\pm0.5\%$ with spinning (top left), and $85.9\%\pm0.6\%$ without spinning (top right). For an unpolarized source, the modulation factor is $0.7\%\pm0.3\%$ with spinning (bottom left), and $2.7\%\pm0.3\%$ without spinning (bottom right).}
  \label{fig:mod_curve}
\end{figure}

Assuming a fully polarized source as bright as RX J1856.5$-$3754 in the presence of a background level mentioned above, we simulated its modulation curves in cases with and without instrument spinning.  The results indicate that a modulation factor of $88.7\%\pm0.5\%$ can be obtained in the case with spinning, and $85.9\%\pm0.6\%$ without spinning. If we assume that the source is unpolarized,  the simulated modulations are $0.7\%\pm0.3\%$ and $2.7\%\pm0.3\%$ for the case with and without instrument spinning, respectively (see Figure ~\ref{fig:mod_curve}).  For these simulations, the exposure time is $10^6$ seconds for the fully polarized source and $10^7$ seconds for unpolarized source. The low degree of modulation for unpolarized sources suggests that we can reach a detection limit of 1\% with instrument spinning, and in most cases when the observing time is less than $10^6$ seconds, the results will be dominated by statistical errors. 

Therefore, the polarimeter has a high degree of response, with a modulation factor close to 0.9, and a low systematics which can be ignored in real observations up to $10^7$ seconds. 

\section{Sensitivity and potential targets}

The sensitivity of a polarimeter is characterized by the minimum detectable polarization (MDP) \cite{Weisskopf2010}. One usually quotes the MDP at 99\% confidence level, 

\begin{equation}
  MDP = \frac{4.29}{\mu R_{\rm S}}\sqrt{\frac{R_{\rm S}+R_{\rm B}}{T}},
  \label{mdp}
\end{equation}
where $\mu$ is the modulation factor, response from a fully polarized source, $R_{\rm S}$ and $R_{\rm B}$ are source and background count rates, respectively, and $T$ is exposure time. This means that if the source is unpolarized, the probability of being detected with a modulation amplitude above this MDP level is 1\%. 

The source flux at 250 eV has been measured with the ROSAT satellite. The source intensity was adopted from the ROSAT all-sky survey bright source catalog \cite{Voges1999}. The band A in the catalog represents an energy band of 0.1--0.4 keV. Therefore, it offers an almost unbiased estimate for the source flux in the LAMP passband.  A constant of $1.02 \times 10^{-2}$ was used to translate the ROSAT band A count rate to LAMP count rate, assuming a total energy-integrated reflectivity of 0.5 eV for unpolarized sources (1.0 eV for s-light and 0 for p-light, see Figure~\ref{fig:ref_curve}), and a detection efficiency of 0.35. The translation is not much dependent on the assumed source spectral shape, as the ROSAT band A is narrow and the LAMP passband sits just in its center.  The background level was estimated to be $1.47 \times 10^{-3}$ counts~s$^{-1}$ on the detector plane according to the raytracing simulation. In Table~\ref{tab:objects}, we listed bright sources of different types (including the XDINS, pulsar, supernova remnant, white dwarf, AGN, black hole X-ray binary, neutron star X-ray binary, blazar, and radio galaxy), their expected count rates and MDPs with an exposure of $10^6$~seconds. 

\begin{center}       
\begin{longtable}{|l|l|l|l|l|} 
\caption{Potential targets for LAMP.} \label{tab:objects} \\
\hline
\rule[-1ex]{0pt}{3.5ex} Type & ROSAT Name & Common Name  & Rate (ct/s) & MDP (\%) \\
\hline
\endhead
\hline \multicolumn{5}{r}{\textit{Continued on next page}} \\
\endfoot
\endlastfoot
\rule[-1ex]{0pt}{3.5ex} XDINS & 1RXS J185635.1-375433 & 1ES 1853-37.9                 & 0.0355 & 2.55 \\
\rule[-1ex]{0pt}{3.5ex} XDINS & 1RXS J072025.1-312554 & 1ES 0718-31.3                 & 0.0152 & 3.94 \\
\rule[-1ex]{0pt}{3.5ex} XDINS & 1RXS J160518.8+324907 & RX J1605.3+3249               & 0.00755 & 5.68 \\
\rule[-1ex]{0pt}{3.5ex} XDINS & 1RXS J080623.0-412233 & RX J0806.4-4123               & 0.00275 & 9.95 \\
\rule[-1ex]{0pt}{3.5ex} XDINS & 1RXS J130848.6+212708 & RX J1308.6+2127               & 0.00175 & 13.1 \\
\rule[-1ex]{0pt}{3.5ex} XDINS & 1RXS J042003.1-502300 & RX J0420.0-5022               & 0.00126 & 16.1 \\
\rule[-1ex]{0pt}{3.5ex} XDINS & 1RXS J214303.7+065419 & RX J2143.0+0654               & 0.00125 & 16.1 \\
\rule[-1ex]{0pt}{3.5ex} Pulsar & 1RXS J065948.1+141431 & PSR B0656+14                  & 0.0156 & 3.88 \\
\rule[-1ex]{0pt}{3.5ex} Pulsar & 1RXS J063354.1+174612 & SN 437                        & 0.00458 & 7.45 \\
\rule[-1ex]{0pt}{3.5ex} Pulsar & 1RXS J105758.6-522703 & PSR B1055-52                  & 0.00376 & 8.32 \\
\rule[-1ex]{0pt}{3.5ex} Pulsar & 1RXS J043714.5-471503 & PSR J0437-47                  & 0.00112 & 17.4 \\
\rule[-1ex]{0pt}{3.5ex} White Dwarf & 1RXS J131621.4+290555 & WD 1314+293 & 0.736 & 0.556 \\
\rule[-1ex]{0pt}{3.5ex} White Dwarf & 1RXS J150209.2+661220 & WD 1501+663                   & 0.184 & 1.11 \\
\rule[-1ex]{0pt}{3.5ex} White Dwarf & 1RXS J200905.6-602537 & WD 2004-605                   & 0.0879 & 1.61 \\
\rule[-1ex]{0pt}{3.5ex} White Dwarf & 1RXS J125702.4+220155 & WD 1254+223 & 0.0613 & 1.93 \\
\rule[-1ex]{0pt}{3.5ex} White Dwarf & 1RXS J215621.5-543820 & WD 2152-548                   & 0.0603 & 1.95 \\
\rule[-1ex]{0pt}{3.5ex} White Dwarf & 1RXS J162909.4+780439 & WD 1631+78                    & 0.0572 & 2 \\
\rule[-1ex]{0pt}{3.5ex} AGN & 1RXS J144207.7+352632 & Mrk 478                       & 0.0499 & 2.14 \\
\rule[-1ex]{0pt}{3.5ex} AGN & 1RXS J123741.4+264229 & IC 3599                       & 0.0422 & 2.34 \\
\rule[-1ex]{0pt}{3.5ex} AGN & 1RXS J042601.6-571202 & 2RE J042537-571348            & 0.0352 & 2.56 \\
\rule[-1ex]{0pt}{3.5ex} AGN & 1RXS J120308.9+443155 & NGC 4051                      & 0.0289 & 2.83 \\
\rule[-1ex]{0pt}{3.5ex} AGN & 1RXS J141759.6+250817 & NGC 5548                      & 0.0279 & 2.88 \\
\rule[-1ex]{0pt}{3.5ex} BHXRB & 1RXS J170248.5-484719 & GX 339-4 & 0.0333 & 2.63 \\
\rule[-1ex]{0pt}{3.5ex} BHXRB & 1RXS J053855.6-640457 & LMC X-3 & 0.0103 & 4.83 \\
\rule[-1ex]{0pt}{3.5ex} BHXRB & 1RXS J195821.9+351156 & Cyg X-1 & 0.00622 & 6.3 \\
\rule[-1ex]{0pt}{3.5ex} BHXRB & 1RXS J053938.8-694515 & LMC X-1 & 0.00104 & 18.2 \\
\rule[-1ex]{0pt}{3.5ex} NSXRB & 1RXS J165749.6+352022 & Her X-1 & 0.131 & 1.32 \\
\rule[-1ex]{0pt}{3.5ex} NSXRB & 1RXS J214429.0+381913 & Cyg X-2 & 0.0374 & 2.48 \\
\rule[-1ex]{0pt}{3.5ex} NSXRB & 1RXS J173143.6-165736 & GX 9+9 & 0.022 & 3.25 \\
\rule[-1ex]{0pt}{3.5ex} NSXRB & 1RXS J182340.5-302137 & 4U 1820-303 & 0.0194 & 3.47 \\
\rule[-1ex]{0pt}{3.5ex} NSXRB & 1RXS J152040.8-571007 & Cir X-1 & 0.00313 & 9.23 \\
\rule[-1ex]{0pt}{3.5ex} SNR & 1RXS J052522.7-655923 & SNR J052529-655905            & 0.0073 & 5.78 \\
\rule[-1ex]{0pt}{3.5ex} SNR & 1RXS J052502.8-693840 & SNR J052501-693842            & 0.00369 & 8.41 \\
\rule[-1ex]{0pt}{3.5ex} SNR & 1RXS J232325.4+584838 & Cas A & 0.00249 & 10.6 \\
\rule[-1ex]{0pt}{3.5ex} Blazar & 1RXS J215852.2-301338 & QSO B2155-304                 & 0.261 & 0.934 \\
\rule[-1ex]{0pt}{3.5ex} Blazar & 1RXS J110427.1+381231 & QSO B1101+384                 & 0.163 & 1.18 \\
\rule[-1ex]{0pt}{3.5ex} Blazar & 1RXS J103118.6+505341 & QSO B1028+511                 & 0.0286 & 2.85 \\
\rule[-1ex]{0pt}{3.5ex} Blazar & 1RXS J113626.6+700932 & QSO B1133+704                 & 0.0276 & 2.9 \\
\rule[-1ex]{0pt}{3.5ex} Blazar & 1RXS J235908.0-303740 & QSO B2356-309                 & 0.0238 & 3.13 \\
\rule[-1ex]{0pt}{3.5ex} Blazar & 1RXS J165352.6+394538 & 4C 39.49                      & 0.0237 & 3.13 \\
\rule[-1ex]{0pt}{3.5ex} Blazar & 1RXS J142832.6+424028 & QSO B1426+428                 & 0.0226 & 3.21 \\
\rule[-1ex]{0pt}{3.5ex} Blazar & 1RXS J111706.3+201410 & 87GB 111429.0+203022          & 0.0223 & 3.23 \\
\rule[-1ex]{0pt}{3.5ex} Blazar & 1RXS J121752.1+300705 & QSO B1215+303                 & 0.0175 & 3.66 \\
\rule[-1ex]{0pt}{3.5ex} Blazar & 1RXS J200925.6-484953 & QSO B2005-489                 & 0.0152 & 3.94 \\
\rule[-1ex]{0pt}{3.5ex} Blazar & 1RXS J101504.3+492604 & QSO B1011+496                 & 0.0136 & 4.17 \\
\rule[-1ex]{0pt}{3.5ex} Blazar & 1RXS J142239.1+580159 & QSO B1422+580                 & 0.0111 & 4.64 \\
\rule[-1ex]{0pt}{3.5ex} Blazar & 1RXS J093037.1+495028 & QSO B0927+500                 & 0.0103 & 4.82 \\
\rule[-1ex]{0pt}{3.5ex} Blazar & 1RXS J232444.9-404053 & 6dFGS gJ232444.7-404050       & 0.0099 & 4.92 \\
\rule[-1ex]{0pt}{3.5ex} Blazar & 1RXS J143917.7+393248 & QSO B1437+398                 & 0.00981 & 4.94 \\
\rule[-1ex]{0pt}{3.5ex} Blazar & 1RXS J032540.8-164607 & 2MASS J03254109-1646169       & 0.00924 & 5.1 \\
\rule[-1ex]{0pt}{3.5ex} Blazar & 1RXS J113630.9+673708 & 2MASX J11363009+6737042       & 0.009 & 5.17 \\
\rule[-1ex]{0pt}{3.5ex} Blazar & 1RXS J124312.5+362743 & 87GB 124048.4+364431          & 0.00896 & 5.19 \\
\rule[-1ex]{0pt}{3.5ex} Blazar & 1RXS J141756.8+254329 & 7C 1415+2556                  & 0.0079 & 5.55 \\
\rule[-1ex]{0pt}{3.5ex} Blazar & 1RXS J122121.7+301041 & QSO B1218+304                 & 0.0077 & 5.62 \\
\rule[-1ex]{0pt}{3.5ex} Radio Galaxy & 1RXS J013741.7+330931 & 3C 48 & 0.0026 & 10.3 \\
\rule[-1ex]{0pt}{3.5ex} Radio Galaxy & 1RXS J022502.9-231254 & QSO B0222-234 & 0.00226 & 11.2 \\
\hline 
\end{longtable}
\end{center}

There are 4 X-ray dim isolated neutron stars, 3 rotation-powered pulsars, 4 supernova remnants,  and a couple dozens of blazars can be detected with an MDP below 10\% with an exposure of $10^6$~seconds.  This list does not include most of the transient X-ray binaries, which are generally missed by the ROSAT all sky survey. 

\section{Conclusion}

Theoretical predictions suggest that X-ray polarimetry will be an important avenue for future high energy astrophysics. The technique is ready for the high-sensitive photoelectric polarimeters, e.g., the XIPE mission \cite{Soffitta2013}, and is also ready for the soft Bragg polarimeters. Thanks to the large collecting area and the focusing capability of multilayer mirrors that lead to high signal to noise ratios, a soft X-ray Bragg polarimeter is still sensitive enough for a few tens of objects even if the passband is ultra narrow.  For these sources, part of the science cannot be accomplished with a photoelectric polarimeter that is only sensitive above 2 keV. For example, polarimetry for the thermal emission from surface of pulsars will allow us to measure the magnetic geometry, test the QED effect, and even identify bare quark stars. These interesting and important sciences can be implemented by a low-cost micro-satellite mission.  


\end{document}